\documentstyle[12pt]{article} \textwidth=17cm \textheight=22.5cm

\def\beq{\begin{equation}} \def\eeq{\end{equation}} \def\bea{\begin{eqnarray}}
\def\eea{\end{eqnarray}}
\def\bq{\begin{quote}} \def\eq{\end{quote}}

\def\um{1/2}
\def\sq{1/\sqrt{2}}

\def\gappeq{\mathrel{\rlap {\raise.5ex\hbox{$>$}} {\lower.5ex\hbox{$\sim$}}}}

\def\lappeq{\mathrel{\rlap{\raise.5ex\hbox{$<$}} {\lower.5ex\hbox{$\sim$}}}}

\begin{document} \pagestyle{empty} \begin{flushright} 
CERN-TH/98-310 \\
DFPD-98/TH/46 \end{flushright}
\vspace*{5mm}
\begin{center} {\bf Models of Neutrino Masses from Oscillations with Maximal Mixing} \\
\vspace*{1cm}  {\bf Guido Altarelli} \\ \vspace{0.3cm} Theoretical Physics Division, CERN \\ CH -
1211 Geneva 23 \\ and
\\Universit\`a di Roma Tre, Rome, Italy \\{\bf
\vspace{0.3cm} Ferruccio Feruglio} \\ \vspace{0.3cm}  
Universit\`a di Padova
\\
and
\\
I.N.F.N., Sezione di Padova, Padua, Italy\vspace{0.3cm} 
\\ \vspace*{2cm}   {\bf Abstract} \\ \end{center}
\vspace*{5mm} \noindent
We discuss models of neutrino masses that lead to a large mixing angle for atmospheric neutrino oscillations. In
particular we study a mechanism where a simple texture for the Dirac matrix leads to the observed pattern of
mixings for a continuous range of Majorana matrices in a reasonably natural way. Both nearly maximal and small
mixing solutions for solar neutrinos are compatible with this model. Possible dynamical realizations are 
discussed
and a detailed example in terms of horizontal U(1) charges is presented.

\vspace*{3cm}  \noindent  

\noindent

\begin{flushleft} CERN-TH/98-310 \\ 
DFPD-98/TH/46 \\
September 1998 \end{flushleft} \vfill\eject 

\setcounter{page}{1} \pagestyle{plain}


Following the experimental results from Superkamiokande \cite{SK} a lot of attention has been devoted to the
problem of a natural explanation of the observed nearly maximal mixing angle for atmospheric
neutrino oscillations. It is possible that also solar neutrino oscillations, if explained by vacuum
oscillations, occur with a large mixing angle \cite{solar}. Large mixing angles are somewhat unexpected because
the observed quark mixings are small and the quark, charged lepton and neutrino mass matrices are to
some extent related in Grand Unified Theories. In a previous paper \cite{AF} we have given a first
discussion of this problem. Within the framework of the see-saw mechanism we focussed on the
interplay between  the neutrino
Dirac and Majorana matrices  which is necessary in order to generate maximal
mixing  \cite{sum}. In the present note we continue the study of this issue. We start by giving a general
formulation of the problem and by specifying our working assumptions. Among our previously proposed strategies
for a natural explanation of maximal mixing, we select and further elaborate on one which we find
particularly plausible and appealing. For our mechanism to work we need that, in analogy with what
is observed for quark and leptons, one mass eigenvalue is dominant for neutrinos (for both the Dirac and the
effective light neutrino mass matrices). Also, in zeroth approximation, we need that the neutrino Dirac mass
matrix, in the basis where charged leptons are diagonal, has only two large entries of comparable magnitude. We
then show that either single or double nearly maximal mixing occur automatically for whatever Majorana mass 
matrix
chosen in a large domain of parameter space, without need of fine tuning. We discuss models where these 
conditions
can be realized and give numerical examples that fit the present data. We finally compare ours with other 
possible
mechanisms.

We start by assuming only three
flavours of neutrinos that receive masses from the see-saw mechanism \cite {ss} and allow all
possible hierarchical patterns for the neutrino mass eigenvalues $m_i$. For reasons of simplicity, we
consider the simplest version of the see-saw mechanism with one Dirac, $m_D$, and one Majorana, $M$,
mass matrix, related to the neutrino mass matrix
$m_{\nu}$, in the basis where the charged lepton mass matrix is diagonal, by
\beq 
m_{\nu}=m^T_D M^{-1}m_D~~~~~.
\label{mmm}
\eeq 
As well known this is not the most general see-saw mechanism because we are not including the
left-left Majorana mass block.
Maximal atmospheric neutrino mixing and the requirement that the electron neutrino does not
participate in the atmospheric oscillations, as indicated by the Superkamiokande \cite{SK} and Chooz
\cite{Chooz} data, lead directly to the following structure of the
$U_{fi}$ (f=e,$\mu$,$\tau$, i=1,2,3) real orthogonal  mixing matrix, apart from sign convention
redefinitions (here we are not interested in CP violation effects: all matrices are taken real)
\beq 
U_{fi}= 
\left[\matrix{
c&-s&0 \cr
s/\sqrt{2}&c/\sqrt{2}&-1/\sqrt{2}\cr
s/\sqrt{2}&c/\sqrt{2}&+1/\sqrt{2}     } 
\right ]~~~~~.
\label{ufi}
\eeq 
This result is obtained by a simple generalization of the analysis of ref. \cite{bar} (also discussed
in ref. \cite{Bal}) to the case
of arbitrary solar mixing angle ($s\equiv\sin{\theta_{sun}}$, $c\equiv\cos{\theta_{sun}}$):
$c=s=1/\sqrt{2}$ for maximal solar mixing (e.g. for vacuum oscillations $\sin^2{2\theta_{sun}}\sim 0.75$) , while
$\sin^2{2\theta_{sun}}\sim 4s^2\sim 5.5\cdot 10^{-3}$ for
the small angle MSW \cite{MSW} solution. The vanishing of
$U_{e3}$ guarantees that 
$\nu_e$ does not participate in the atmospheric oscillations and the relation
$|U_{\mu3}|=|U_{\tau3}|=1/\sqrt{2}$ implies maximal mixing for atmospheric neutrinos. Note that we
are assuming only two frequencies, given by 
\beq\Delta_{sun}\propto m^2_2-m^2_1,~~~~~~~
\Delta_{atm}\propto m^2_3-m^2_{1,2}\label{fre}
\eeq
The effective light
neutrino mass matrix is given by
$m_\nu=Um_{diag}U^T$ with
$m_{diag}=Diag[m_1,m_2,m_3]$. For generic $s$ one finds
\beq 
m_{\nu}= \left[\matrix{
2\epsilon&\delta&\delta\cr
\delta&\frac{m_3}{2}+\epsilon_2&-\frac{m_3}{2}+\epsilon_2\cr
\delta&-\frac{m_3}{2}+\epsilon_2&\frac{m_3}{2}+\epsilon_2     } 
\right]~~~~~, 
\label{mnu}    
\eeq 
with 
\beq
\epsilon=(m_1 c^2+m_2 s^2)/2~~,~~~~~\delta=(m_1-m_2) c s/\sqrt{2}~~,
~~~~~\epsilon_2=(m_1 s^2+m_2 c^2)/2~~~~~.
\label{ede}    
\eeq
With respect to the notation of ref. \cite{AF} we have reinstated the normalization
factor $m_3/2$ in $m_{\nu}$. We see that the existence of one maximal mixing and
$U_{e3}=0$ are the most important input that leads to the matrix form in eq. (\ref{mnu},\ref{ede}). The value of
the solar neutrino mixing angle can be left free. While the simple parametrization of the matrix U in
eq. (\ref{ufi}) is quite useful to guide the search for a realistic pattern of neutrino mass
matrices, it should not be taken too literally. In particular the data do not exclude a
non-vanishing $U_{e3}$ element. In most of the Superkamiokande allowed region the bound by Chooz
\cite{Chooz} amounts to  $|U_{e3}|\lappeq 0.2$. In the region not covered by Chooz $|U_{e3}|$ can
even be larger \cite{fogli,hall}. Thus neglecting
$|U_{e3}|$ with respect to
$s$ in eq. (\ref{ufi}) is not really justified. Also note that in presence of a large hierarchy
$|m_3|\gg |m_{1,2}|$ the  effect of neglected parameters
in  eq. (\ref{mnu}) can be enhanced by $m_3/m_{1,2}$ and produce seizable corrections. A non
vanishing
$U_{e3}$ term can lead to different $(m_\nu)_{12}$ and  $(m_\nu)_{13}$ terms. Similarly a deviation
from maximal mixing $U_{\mu 3}\not=U_{\tau 3}$ distorts the $\epsilon_2$ terms in the 23 sector of
$m_{\nu}$. Therefore, especially for a large hierarchy, there is more freedom in the small terms in order to
construct a model that fits the data than it is apparent from eq. (\ref{mnu}).

Given the observed frequencies and our notation in eq. (\ref{fre}), there are three possible hierarchies of mass
eigenvalues:
\bea
	{\rm A}& : & |m_3| >> |m_{2,1}| \nonumber\\
	{\rm B}& : & |m_1|\sim |m_2| >> |m_3| \nonumber\\
	{\rm C}& : & |m_1|\sim |m_2| \sim |m_3|
\label{abc}
\eea
(in case A there is no prejudice on the $m_1$, $m_2$ relation). For B and C different subcases are
then generated according to the relative sign assignments for $m_{1,2,3}$. For each case we can set to
zero the small masses and mixing angles and find the effective light neutrino matrices which are
obtained both for double and single maximal mixing. Note that here we are working in the
basis where the charged lepton masses are diagonal, and approximately given by
$m_l=Diag[0,0,m_{\tau}]$. For model building one has to arrange both the charged lepton and the
neutrino mass matrices so that the neutrino results coincide with those given here after
diagonalization of charged leptons. For example, in case A,
$m_{diag}=Diag[0,0,m_3]$ and we obtain
\beq
m_{\nu}/m_3= \left[\matrix{
0&0&0\cr
0&\frac{1}{2}&-\frac{1}{2}\cr
0&-\frac{1}{2}&\frac{1}{2}    } \right]~~~~~. 
\label{a}
\eeq
In this particular case the results are the same for double and
single maximal mixing. Note that the signs correspond to the phase convention adopted in
eq. (\ref{ufi}). If one prefers all signs to be positive it is sufficient to invert the sign of the
third row of the matrix U in eq. (\ref{ufi}). We can similarly proceed in the other cases and we obtain
the results in table I (where the overall mass scale was dropped), which we now discuss.
\\[0.1cm]
{\begin{center}
\footnotesize
\begin{tabular}{|c|c|c|c|}   
\hline                        
& & double & single \\
& $m_{diag}$ & maximal & maximal \\
& & mixing & mixing \\
\hline
& & & \\
A & Diag[0,0,1] & 
$\left[
\matrix{
0&0&0\cr
0&\um&-\um\cr
0&-\um&\um}
\right]$ &
$\left[
\matrix{
0&0&0\cr
0&\um&-\um\cr
0&-\um&\um}
\right]$ \\
& & & \\
\hline
& & & \\
B1 & Diag[1,-1,0] & 
$\left[
\matrix{
0&\sq&\sq\cr
\sq&0&0\cr
\sq&0&0}
\right]$ &
$\left[
\matrix{
1&0&0\cr
0&-\um&-\um\cr
0&-\um&-\um}
\right]$ \\
& & & \\
\hline
& & & \\
B2 & Diag[1,1,0] &
$\left[
\matrix{
1&0&0\cr
0&\um&\um\cr
0&\um&\um}
\right]$ &
$\left[
\matrix{
1&0&0\cr
0&\um&\um\cr
0&\um&\um}
\right]$ \\
& & & \\
\hline
& & & \\
C1 & Diag[-1,1,1] & 
$\left[
\matrix{
0&-\sq&-\sq\cr
-\sq&\um&-\um\cr
-\sq&-\um&\um}
\right]$ &
$\left[
\matrix{
-1&0&0\cr
0&1&0\cr
0&0&1}
\right]$ \\
& & & \\
\hline
& & & \\
C2 & Diag[1,-1,1] & 
$\left[
\matrix{
0&\sq&\sq\cr
\sq&\um&-\um\cr
\sq&-\um&\um}
\right]$ &
$\left[
\matrix{
1&0&0\cr
0&0&-1\cr
0&-1&0}
\right]$ \\
& & & \\
\hline
& & & \\
C3 & Diag[1,1,-1] & 
$\left[
\matrix{
1&0&0\cr
0&0&1\cr
0&1&0}
\right]$ &
$\left[
\matrix{
1&0&0\cr
0&0&1\cr
0&1&0}
\right]$ \\
& & & \\
\hline                        
\end{tabular}                             
\end{center}}
\vspace{3mm} 
{\bf Table~I} : Zeroth order form of the neutrino mass matrix 
for double and single maximal mixing, according to the
different possible hierarchies given in eq. (\ref{abc}). 
\vspace{0.5cm}  

The completely degenerate case C is the only one that could in principle accommodate neutrinos as hot
dark matter together with solar and atmospheric neutrino oscillations. For this the common mass should be
around 1-3 eV. Then the solar frequency could be given by a small 1-2 splitting, while the atmospheric
frequency could be given by a still small but much larger 1,2-3 splitting. Note that we have not included in 
Table
1 the degenerate case C with three equal signs. Clearly the mixing matrix in this case is completely determined 
by
the small degeneracy splitting corrections and the zeroth order approximation is irrelevant. A
strong constraint arises in the degenerate case from neutrinoless double beta decay which requires that the
$ee$ entry of
$m_{\nu}$ must obey
$|(m_{\nu})_{ee}|\leq 0.46~{\rm eV}$ \cite{dbeta}. As observed in ref. \cite{GG}, this bound can only be 
satisfied if bimixing (that is
double maximal mixing) is realized. In fact we see from the expression of $\epsilon$ in eq. (\ref{ede})
that we need $m_1\sim -m_2$ and
$c^2\sim s^2$ for a cancellation to occur in $(m_{\nu})_{ee}$. Since the solar mixing can only be either near
maximal or very small, only the bimixing solutions C1 and C2 survive (which are physically equivalent
\footnote{The solutions C1 and C2 are related by the exchange $s\leftrightarrow c$.}), as can
be verified in table 1. So all other solutions of type C can only survive if the common mass is below $0.46~
{\rm eV}$.
We think that it is not at all clear at the moment that a hot dark matter component is really
needed \cite{kra}. However the main reason to consider the fully degenerate solution is 
that it is compatible
with hot dark matter, so that only the solution C1/C2 in the case of bimixing is interesting.
Note that for degenerate masses with $m\sim 1-3~{\rm eV}$ we need a relative splitting $\Delta m/m\sim
\Delta m^2_{atm}/2m^2\sim 10^{-3}-10^{-4}$ and a much smaller one for solar neutrinos explained by vacuum
oscillations:
$\Delta m/m\sim 10^{-10}-10^{-11}$. We are unable
to imagine a natural mechanism compatible with unification and the see-saw mechanism leading to such a
precise near symmetry. An ansatz of this sort has been proposed in ref. \cite{fri}. 

Of the remaining possibilities two are particularly remarkable for their simplicity: solution A
(eq. (\ref{a})) and solution B1 ( for bimixing). Of the six independent entries of a symmetric matrix
three (four) are zero for solution A (B1 bimixing). However, solution A requires the entries in the
23 sector to be related (in particular the 23 subdeterminant must be zero, in order for two of the
eigenvalues to be vanishing in zeroth order, as assumed in this case), solution B1 requires the two non zero
entries in the first row or column to be nearly equal. In the following we shall discuss how these patterns can
be generated in a natural way.

As a first orientation we observe that, after diagonalization of the charged lepton Dirac mass matrix, we still
have the freedom of a change of basis for the right-handed neutrino fields. In fact right-handed charged lepton and
neutrino fields, as opposed to left-handed fields, are uncorrelated by $SU(2)\bigotimes U(1)$ gauge symmetry. We
can use this freedom to make the Majorana matrix diagonal: $M^{-1}=O^Td_MO$ with $d_M=Diag[1/r_1,1/r_2,1/r_3]$.
Then if we parametrize the matrix $Om_D$ by $z_{ab}$ we have:
\beq
m_{\nu ab}=(m_D^TM^{-1}m_D)_{ab}=\sum_c \frac{z_{ca}z_{cb}}{r_c}.\label{cc}
\eeq From this expression we see that, while we can always arrange the twelve parameters $z_{ab}$ and $r_a$
to arbitrarily fix the six independent matrix elements of $m_{\nu}$, case  A is special in that
it can be approximately reproduced in two particularly simple ways, without relying on precise
cancellations among different terms: 

\noindent
i) One of the right-handed neutrinos is particularly light and, in first
approximation, it is only coupled to $\mu$ and $\tau$ \cite{king}. 
Thus, $r_c\sim \eta$  (small) and $z_{c1}\sim 0$. In this
case
\cite{hall,hall2} the 23 subdeterminant vanishes, and one only needs the ratio
$|z_{c2}/z_{c3}|$ to be close to 1. 

\noindent
ii) There are only two large entries in the $z$
matrix, $|z_{c2}|\sim |z_{c3}|$, and the three eigenvalues $r_a$ are of comparable magnitude (or, at
least, with a less pronounced hierarchy than for the $z$ matrix elements). Then, again, the
subdeterminant 23 vanishes and one only needs the ratio
$|z_{c2}/z_{c3}|$ to be close to 1. 

The possibility ii), which was noticed in our previous paper
\cite{AF}, involves no more fine tuning than i). One solution is more specific on M, the other on
$m_D$. Note that solution B1 requires a more delicate balance between the Dirac and Majorana matrices, which,
however, can be realized in particular models \cite{hall,hall2}, as we shall see in the following.

We now discuss the mechanism ii) in more detail. We start with the order zero approximation, as in eq.
(\ref{a}) or table 1, where all small entries in the matrix are set to zero. Assume that, in the basis where
charged leptons are diagonal, the Dirac matrix $m_D$, defined by the bilinear ${\bar \psi_R}m_D\psi_L$, 
takes the approximate form:
\beq
m_D\propto 
\left[\matrix{
0&0&0\cr
0&0&0\cr
0&x&1    } 
\right]~~~~~. 
\label{md0}
\eeq
This matrix has the property that for a generic Majorana matrix $M$ one finds:
\beq
m_{\nu}=m^T_D M^{-1}m_D\propto 
\left[\matrix{
0&0&0\cr
0&x^2&x\cr
0&x&1    } 
\right]~~~~~. 
\label{mn0}
\eeq
The only condition on $M^{-1}$ is that the 33 entry is non zero. The reason for this insensitivity to
$M$ is that $m_D$ given by eq. (\ref{md0}) can be diagonalized by a transformation of the form $V^T m_D
U=d\propto Diag[0,0,1]$ with $V$ (the right-handed field transformation) given by a block-diagonal
matrix with $(V)_{33}=1$.
Then we can write $m_{\nu}=U d V^T M^{-1} V d U^T$. 
The diagonal matrix $d$ acts like a projector so that only the 33 entry of
$d V^T M^{-1} V d$ is non vanishing. Here it is crucial that the neutrino eigenvalues are approximately
$(0,0,m_3)$. Finally the $U$ transformation (the left-handed field rotation) brings this matrix 
into the
form of eq. (\ref{mn0}) that for $|x|->1$ approaches that required by solution A. Note that the 23
subdeterminant of $m_{\nu}$ is vanishing for all values of $x$. The matrix $U$ is directly the neutrino
mixing matrix. Its form is:
\beq
U=  \left[
\matrix{
c&-s&0\cr
sc_{\gamma}&cc_{\gamma}&-s_{\gamma}\cr
ss_{\gamma}&cs_{\gamma}&c_{\gamma}  } 
\right]~~~~~, 
\label{uu}
\eeq
with 
\beq
s_{\gamma}=-x/r~~,~~~~~c_{\gamma}=1/r~~,~~~~~r=\sqrt{1+x^2}~~~~~.
\label{scr}
\eeq
We see that at $|x|\not=1$, $U_{e3}$ still vanishes but the atmospheric neutrino mixing is no more
maximal. Rather one has 
\beq
\sin^2{2\theta}=4s^2_{\gamma}c^2_{\gamma}=\frac{4 x^2}{(1+x^2)^2}~~~~~.
\label{sin}
\eeq
Thus the bound $\sin^2{2\theta}\gappeq0.8$ translates into  $0.6\lappeq |x|\lappeq 1.6$. It is interesting to
recall that in ref.\cite{ellis} it was shown that the mixing angle can be amplified by the running from a
large mass scale down to low energy.

A zeroth order texture as in eq. (\ref{md0}) with generic entries $(m_D)_{23}=A$ and $(m_D)_{33}=B$ 
can well
apply for both charged leptons and neutrinos. If for charged leptons we start
with $A_l$ and $B_l$ and for neutrinos with $A_{\nu}$ and $B_{\nu}$, after diagonalisation of the
charged leptons by a left-handed matrix of the form in eq. (\ref{uu}) with $s=0$, one obtains a matrix
of the same form with
$A/B=x$, as required for
$m_D$ in eq. (\ref{md0}). Precisely the value of $|x|$ is given by $|x|=|(A_lB_{\nu}-
B_lA_{\nu})/(A_lA_{\nu}+B_lB_{\nu})|$.

It remains for us to discuss if the proposed strategy is not completely ruined by going beyond the
zeroth order approximation. In fact terms of order $\lambda$, $\lambda^2$ ... have to be added to
$m_D$ to introduce a hierarchical structure as observed for quarks and leptons. Since, in general, also
M will be hierarchical in terms of some other small parameter $\eta$, $M^{-1}$ can contain terms of
order $1/\eta$, $1/\eta^2$..... In the see-saw product, the ratios of a power of $\lambda$ and a
power of $\eta$ can generate unwanted additional terms of order one or even larger that can spoil the
zeroth order result. Obviously a necessary condition is that the Majorana hierarchy is not too
strong in comparison of the Dirac hierarchy, or that $\eta$ is not smaller than some power of
$\lambda$.
We discuss this issue by showing that reasonable models exist where the mechanism is indeed realised.

An attractive approach to reproduce the observed features of quark and lepton mass matrices is
in terms of horizontal symmetries, in particular, for simplicity, U(1) charges \cite{fro}. When the charges 
do not match by $n$
the corresponding mass term is suppressed by $n$ powers of a small parameter that arises from the vacuum 
expectation
value (divided by some large mass) of one or more extra scalar fields $\theta_i$ that carry one unit of charge.
There can be two different fields with opposite charges $\theta_\pm$
\footnote{We regard the conjugate field $\theta^*$ as distinct from $\theta$. Therefore
$\theta_-$ may coincide with $(\theta_+)^*$.}, so that the suppression factors can involve
two different small parameters, according to the sign of $n$. We now show that our mechanism can be realised in a
simple U(1) model with two oppositely charged fields $\theta_\pm$. 


We want to construct an example of charge assignments that lead to the desired form for the Dirac and
Majorana matrices. In particular, we demand that, for the Majorana matrix M, the determinant is of order 1
(because we want to avoid that one lightest eigenvalue exists, which would be typical of mechanism i), while our
case prefers comparable eigenvalues) and the same is true for the matrix element $(M^{-1})_{33}$ (which plays a
special role in our mechanism). One possibility is to start with the following charges for the lepton doublet L
and the right-handed neutrino fields:
\beq
L=(2,0,0);~~~~~~~~\bar R_{\nu}=(1,-1,0)\label{ch}
\eeq
which lead to the textures:
\beq
m_D\propto \left[\matrix{
\lambda^3&\lambda&\lambda\cr
\lambda&\lambda^\prime&\lambda^\prime\cr
\lambda^2&1&1    } 
\right];~~~~~
M\propto \left[\matrix{
\lambda^2&1&\lambda\cr
1&\lambda^{\prime2}&\lambda^\prime\cr
\lambda&\lambda^\prime& 1    } 
\right]~~~~~. 
\label{m1}
\eeq
It is simple to verify that, for generic coefficients, the effective light neutrino matrix is of the form:
\beq
m_\nu\propto\left[\matrix{
\lambda^4&\lambda^2&\lambda^2\cr
\lambda^2&1&1\cr
\lambda^2&1&1} \right]~~~~~, 
\label{mn1}
\eeq
However, the crucial property is that the 23 subdeterminant of $m_\nu$ automatically vanishes in the limit
$\lambda\rightarrow 0$, because of the simple mathematical property displayed by eqs. (\ref{md0}), (\ref{mn0}).
Actually, for generic coefficients, the 23 minor is of order $\lambda\lambda^\prime$. We now specify the charges of
right-handed charged leptons. The requirements are that the hierarchy of eigenvalues corresponds to the  observed
e, $\mu$, $\tau$ mass ratios and that after diagonalisation of the charged leptons the form of the Dirac matrix for
neutrinos is not spoiled (for this we need that the left-handed diagonalising matrix is of the form in
eq. (\ref{uu}) with $s$ small). An acceptable choice for the right-handed lepton charges is:
\beq
\bar R_l=(-4,1,0)\label{chl}
\eeq
which leads to:
\beq
m_l\propto\left[\matrix{
\lambda^{\prime2}&\lambda^{\prime4}&\lambda^{\prime4}\cr
\lambda^3&\lambda&\lambda\cr
\lambda^2&1&1} \right]~~~~~. 
\label{ml1}
\eeq
This matrix has eigenvalues of order $(\lambda^{\prime2},\lambda,1)$. For $\lambda \sim \lambda^\prime \sim
10^{-1}-10^{-2}$ and generic coefficients one obtains numerically reasonable solutions. In this analysis 
we have also taken into account the most general kinetic terms
compatible with the chosen charge assignement. We found that
numerically acceptable solutions are generated even in the presence of these terms.
(See also ref. \cite{lns}). Most of them are of the
bimixing type. This is because, as seen from eq. (\ref{mn1}), 
the 11 entry of $m_\nu$ is much smaller than the 12,
13 entries. Given the general expression eq. (\ref{mnu}) this particular pattern corresponds to double nearly
maximal mixing.

It is interesting to note that 
if all the $U(1)$ charges for the left-handed lepton doublets and the right-handed anti-neutrinos have the 
same sign, the relation between the Dirac and
Majorana hierarchies is too tight for the mechanism to work. In fact in this case, the neutrino Dirac
and Majorana mass matrices can be written as \cite{ram} $m^{\nu}_D=Q_{\nu \bar R}Y_D Q_L$ and
$M=Q_{\nu \bar R}Y_M Q_{\nu \bar R}$ where the $Q$'s are diagonal matrices of the form
$Q=Diag(\lambda^{q_1},\lambda^{q_2},\lambda^{q_3})$ in terms of one single small parameter $\lambda$ and the
charges $q_i$. $Y$ are generic non hierarchical Yukawa matrices. It is easy to see that in constructing
$m_{\nu}$ the right-handed charges drop away and $m_{\nu}=Q_LY^T_DY^{-1}_MY_D Q_L$. With the above
left-handed charge assignments the resulting matrix $m_{\nu}$ indeed has the 22, 23, 32, 33 entries of order 1
while all others are suppressed, but the 23 subdeterminant is not vanishing because the matrix
$Y^T_DY^{-1}_MY_D$ is completely generic. Thus the resulting eigenvalue structure is not the right one
in general and must be fixed by hand. It is easy to see why our  mechanism does not work in this
case. In fact, the same reason that ensures the cancellation of right-handed charges in the see-saw product
makes the connection between the hierarchies of the Dirac and Majorana matrices such that the $\lambda$ terms of
$m_D$ and the $1/\lambda$ terms of $M^{-1}$ necessarily generate finite terms that spoil the vanishing of the
23 subdeterminant. So models of type A cannot in general be explained in terms of 
all positive (or all negative) $L$ and ${\bar R}_\nu$ charges.
For our mechanism to work we need that the Dirac and Majorana mass matrices are generated
by less tightly connected terms, as in the above example.
This could also be the case in models based on GUT's \cite{Large}, for
example Susy $SO(10)$ or flipped $SU(5)$, where $m_D$ and M are generated by different Higgs multiplets and
representations.  Note that the model B1 (bimixing) can be reproduced in terms of left-handed charges (a,-a,-a) for
$(e_L,\mu_L,\tau_L)$, as observed in ref. \cite{hall,hall2}, but the 12 and 23 entries, of order 1, must be 
arranged to be nearly equal.  

In conclusion, we have studied a mechanism, introduced in \cite{AF}, that can explain the observed pattern of
neutrino oscillations, both for double and single maximal mixing. In our case both for Dirac and light
neutrino masses there is one dominant eigenvalue: $|m_3|>>|m_{1,2}|$ as in case A of table 1. Our proposed
mechanism only contains a minimum of fine tuning and compares well with all other mechanisms sofar presented
\cite{Large,sterili,Rbreaking,abelian,large}. Some
of them were also discussed here. From these examples we learn that after all large mixings are not so 
unplausible
as it appears at first sight. In all of these models there is no reason why the atmospheric neutrino mixing 
should
be maximal but it can be large. In many cases once the atmospheric mixing is large also the solar mixing turns 
out
to be large, but this is not necessarily true in all models, for example in our model.

\vfill
\end{document}